\documentstyle[aas2pp4]{article}
\def\refindent{\par\noindent\hangindent=3pc\hangafter=1 }

\def\aasup#1#2#3{\refindent#1, A\&AS, #2, #3}

\def\apj#1#2#3{\refindent#1, {\it Ap. J.}, {\bf#2}, #3.}

\def\mnras#1#2#3{\refindent#1, {\it M.N.R.A.S.}, {\bf#2}, #3.}

\def\>{$>$}
\def\<{$<$}

\def\simlt{\lower.5ex\hbox{$\; \buildrel < \over \sim \;$}}
\def\simgt{\lower.5ex\hbox{$\; \buildrel > \over \sim \;$}}
\def\sqr#1#2{{\vcenter{\hrule height.#2pt
      \hbox{\vrule width.#2pt height#1pt \kern#1pt
         \vrule width.#2pt}
      \hrule height.#2pt}}}

\begin{document}
\centerline{Submitted to the Editor of the Astrophysical Journal Letters}
\centerline{Revised May 28, 1997}
\vskip 0.3in
\title{Magnetic Flares and the Observed $\tau_T \sim 1$ in
  Seyfert Galaxies}

\author{Sergei Nayakshin$^*$ and Fulvio Melia$^{*\dag}$\altaffilmark{1}}
\affil{$^*$Physics Department, University of Arizona, Tucson, AZ 85721}
\affil{$^{\dag}$Steward Observatory, University of Arizona, Tucson, AZ 85721}

\altaffiltext{1}{Presidential Young Investigator.}


\begin{abstract}
We here consider the pressure equilibrium during an intense magnetic flare
above the surface of a cold accretion disk.  Under the assumption that
the heating source for the plasma trapped within the flaring region is 
an influx of energy transported inwards with a group velocity close to $c$,
e.g., by magnetohydrodynamic waves, this pressure equilibrium can constrain 
the Thomson optical depth $\tau_T$ to be of order unity. 
We suggest that this may be the reason why $\tau_T\sim 1$ in Seyfert Galaxies. 
We also consider whether current data can distinguish between the spectrum
produced by a single X-ray emitting region with $\tau_T\sim 1$ and that
formed by many different flares spanning a range of $\tau_T$. We find 
that the current observations do not yet have the required energy
resolution to permit such a differentiation.  Thus, it is possible that 
the entire X-ray/$\gamma$-ray spectrum of Seyfert Galaxies is produced
by many independent magnetic flares with an optical depth $0.5<\tau_T<2$. 
\end{abstract}




%
\section{Introduction}

Although the X-ray emission from Seyfert galaxies is still not fully
understood, there are indications that the mystery is starting to
unfold.  It is now believed that the emission mechanism is
most likely thermal (see, e.g., Svensson 1996; Zdziarski et al. 1996).
A flattening of the spectrum above about 10 keV (e.g., Nandra \& 
Pounds 1994) strongly suggests the presence of a component due to
the reflection of the intrinsic X-ray spectrum by nearby cold matter,
which means that the X-ray emitting region is probably located above 
the accretion disk. In addition, the fact that the UV and X-ray 
luminosities from these sources are comparable and that their
time variations are tightly correlated suggests that the UV
and X-ray emitting regions are located very close to each other.
There are also indications that the broad and redshifted Fe
K$\alpha$ line originates within the cold accretion disk which
extends inwards to the last stable orbit (e.g., Iwasawa et al. 1996). 

Many of these observational characteristics fit within the currently
popular two-phase patchy accretion disk-corona model
(Haardt \& Maraschi 1991; Haardt \& Maraschi 1993; Haardt et al. 1994), 
in which the X-ray emitting region is located within `active
regions' (AR) above the disk, where the energy dissipation rate greatly 
exceeds the local gravitational dissipation inside the disk---at least for the 
lifetime of the region. Recent calculations by Zdziarski et al. (1996)
show that the Thomson optical depth $\tau_T$ in Seyfert galaxies must be close
to unity.  From a theoretical perspective, this is highly significant
given that the sample of sources covers a broad range of black hole
masses, accretion rates, and other system parameters. Thus, it is
not easy to understand why $\tau_T\sim 1$, especially since the energy
transport mechanism has not yet been identified.

Haardt et al. (1994) suggested that the ARs may be magnetic flares
occurring above the accretion disk's atmosphere and showed that their
compactness $l$ may be quite high ($\sim 30$), so that pairs can be created.
The local compactness parameter $l$ is defined as 
$l\equiv L\sigma_T/R m_e c^2$, where $L$ is the source 
luminosity, $R$ is the source size and $\sigma_T$ is the Thomson 
cross section. In principle, it is possible to obtain
still larger values for the compactness parameter ($l\sim$ few hundred;
see Nayakshin \& Melia 1997a), thus creating enough pairs to account for the 
observed $\tau_T\simeq 1.0^{+0.4}_{-0.2}$ (Zdziarski et al. 1996). 
This explanation for the observed value of $\tau_T\sim 1$ based on
the pair equilibrium condition, however, relies on the assumption
that the particles are confined to a rigid box, so that no pressure constraints
need to be imposed. This is unphysical for a magnetic flare where the 
particles are free to move along the magnetic field lines.
Therefore, as far as the two-phase model {\it without} a proper pressure equilibrium
condition is concerned, the Thomson optical depth is a parameter, rather than
a calculable quantity.

In this {\it Letter}, we attempt to address this question of pressure
equilibrium during an energetic magnetic flare above the surface of a cold
accretion disk. We show that under the assumptions (i) that the
energy is supplied to the X-ray emitting region by magnetohydrodynamic waves with
a group velocity close to the speed of light and (ii) that the emitting region
has a very large compactness parameter $l\gg 1$, the pressure equilibrium
between the outwardly directed X-ray flux and the inwardly directed
MHD waves leads to the requirement that $\tau_T \sim 1$-$2$, in
accordance with the apparently ``universal'' value of $\tau_T$ seen in Seyfert galaxies.
We will also show that both assumptions are likely to be valid during
a physically plausible magnetic flare above the surface of a cold accretion disk.

\section{The Energy Supply Mechanism}

The `universal' X-ray spectral index of Seyfert Galaxies (e.g., Nandra \&
Pounds 1994) suggests that the emission mechanism is thermal Comptonization
with a $y$-parameter close to one (Haardt \& Maraschi 1991; 
Haardt \& Maraschi 1993; Fabian 1994). For the emission to be
dominated by Comptonization, the compactness parameter needs
to be large ($l\gg 1$; e.g., Fabian 1994).  Observations of Seyfert 
Galaxies point to a global compactness parameter $\sim
1-100$ (Svensson 1996 and references cited therein). 

For electrons and protons at a single temperature $T_e$
and an electron number density $n_e$ with the assumption of
neutrality, the gas pressure is $2 n_e kT$.  The radiation 
energy density may be recast in terms of the luminosity $L$ 
of the source, and therefore its compactness $l$.
The total pressure is
\begin{equation}
P = {m_e c^2\over \sigma_T R}\,\left [ 2\tau_T \,\Theta_e +
l (1+\tau_T)/3\,\right ]{\rm ,} 
\end{equation}
where $\Theta_e\equiv k T_e/m_e c^2$ is the dimensionless 
electron temperature. The Compton $y$-parameter for thermal electrons is 
$y= 4\Theta_e \tau_T (1+\tau_T)(1+4\Theta_e)$
(see Rybicki \& Lightman 1979) and is of order 1. Thus,
since the dimensionless gas pressure in Equation (1) is always 
smaller than $y$, the radiation pressure dominates over the
gas pressure in a one-temperature plasma when $l\gg 1$. 

One consequence of this is that the amount of energy escaping
from the source even during one light crossing time is larger 
than the total particle thermal energy. Thus, there must be an 
agent that energizes the particles to enable them to radiate 
at this high rate, and the presence of this agent must 
be dynamically consistent with the state of the system.
We foresee two possibilities for the nature of this `agent': 
(i) the gravitational field, and (ii) an external flow
of energy into the system. These two cases are quite distinct physically.

Insofar as the first possibility is concerned, the gravitational
potential energy of the plasma (primarily that of the protons)
is dissipated as the gas sinks deeper into the well of the
black hole.  The gravitational field does not provide a pressure, but
it does compress the gas.  However, this leads to an internal (radiation
plus gas) pressure that varies from source to source as the physical
conditions change.  There does not appear to be a scale that sets
$\tau_T$ to have a value of $1$.  For example, in standard accretion disk
theory, the inner radiation pressure-dominated regime has an optical depth
that depends on several parameters, such as the accretion rate
and the $\alpha$-parameter (Shakura \& Sunyaev 1973).  The $\alpha$-parameter
reflects the rate at which the protons `use up' their potential energy,
and so a change in this rate leads to a change in the equilibrium optical depth.
It is even less obvious why $\tau_T$ should be $\sim 1$ in the gas pressure dominated
regimes since there the pressure has no reference to $\tau_T$ at all.
It seems that when the pressure equilibrium is dictated by the 
gravitational field (e.g., due to a compression of the X-ray emitting
region), the Thomson optical depth should span a range of 
values depending on the source geometry, the specific parameter values
and the particle interactions assumed to operate in the source.

This is not so when the energy is supplied to the X-ray emitting region by 
an inflow of energy, e.g., via a magnetic field.
The principal difference between the two cases is
that the dynamic portion of the magnetic field supplies a ``ram'' pressure
that is related in a known way to its energy density. If the magnetic
energy flux into the X-ray emitting region is known, this also constrains
the inwardly directed momentum flux (the compressional force) into the 
system.  Thus, the compressional force exerted on the active region by
the magnetic field is expected to correlate with the source luminosity.
What makes this useful in terms of setting the optical depth of the system is
that a similar correlation exists between the luminosity and the outwardly
directed radiation pressure in the emitting region.  But in this case,
the pressure also depends on $\tau_T$.  Assuming a spherical geometry
for simplicity, the radiation pressure is $P_r \simeq \tau_T F_r/c$,
where $L\approx 4\pi R^2 F_r$ in terms of the source radius $R$ and radiation
flux $F_r$.  Thus, since all the balance equations are to first order
linear in $F_r$, it is anticipated that the pressure and energy
equilibria of the system point to a unique value of $\tau_T$.
We explore this possibility in the next section.

\section{Pressure Equilibrium For Externally Fed Sources}

Let us first suppose that the X-ray source is a sphere with Thomson optical depth
$\tau_T$, and that the energy is supplied radially by magnetohydrodynamic waves.
The waves carry an energy density $\varepsilon$ and propagate with velocity $v_a$. 
For definitiveness, we assume that these are Alfv\'en waves, in which case the 
momentum flux that enters the X-ray source is $(1/2)\,\varepsilon$.
The magnetic energy of the Alfv\'en waves is in equipartition with
the oscillating part of the particle energy density, and so we can estimate
the gas pressure as being of the same order as the ram pressure of the
oscillating part of the magnetic field, i.e. $(1/2)\,\varepsilon$.
Finally, we assume that all of the wave energy and momentum are absorbed by 
the source.

The energy equilibrium for the AR is then given by 
\begin{equation}
F_r = \varepsilon v_a\;,
\end{equation}
whereas in pressure equilibrium 
\begin{equation}
P_r\simeq \tau_T F_r/c \simeq \varepsilon\;.
\end{equation}
Dividing the latter equation by the former, one obtains for the equilibrium
Thomson optical depth:
\begin{equation}
\tau_T \simeq {c\over v_a}
\end{equation}
This value does not depend on luminosity, but it does of course depend on
the geometry and $v_a$. Nayakshin \& Melia (1997a) show that the 
Alfv\'en velocity is close to $c$ for typical conditions in an accretion
disk if its luminosity is a sizable fraction of the Eddington luminosity 
and the magnetic field strength in the AR is comparable to the disk thermal energy
density.

Suppose now that the geometry is not perfectly
spherically symmetric, and that instead the Alfv\'en waves can enter the 
X-ray source through an area $A_a$, but the radiation 
leaves through an area $A_r \ga A_a$, which is plausibly just the total
area of the AR. This situation may occur if part of the X-ray source is
confined by other than the Alfv\'en wave ram pressure, e.g., by
the underlying (non-dynamic) large-scale magnetic field (see below). In this
case, since the energy balance is now $F_r A_r = \varepsilon v_a A_a$,
the equilibrium $\tau_T$ is changed to
\begin{equation}
\tau_T \simeq \left({c\over v_a}\right)
\left({A_r\over A_a}\right)\;.
\end{equation}

To understand the scale represented by the bracketed quantities in
this equation, let us consider the physical conditions that are likely to be
attained during a short-lived and very energetic magnetic flare
above the standard $\alpha$-disk.  The magnetic field energy density is 
a fraction of the underlying disk energy density and the typical 
size $\Delta R_a$ of the flare is expected to be of the order of the 
disk scale height (Galeev et al. 1979; Haardt et al. 1994; 
Nayakshin \& Melia 1997a).  Now, the confinement of the
plasma inside the flare, and the observed condition $l\gg 1$, require that
$B^2/8\pi \gg P_r \gg P_g$. Since the magnetic stress is much larger than
any other stress, the magnetic flux tube adjusts to be in a stress-free vacuum 
configuration. The tube is thick (meaning that its cross sectional radius is 
of the order of its length), since the pressure in the disk's atmosphere 
is insufficient to balance the tube's magnetic field pressure.  This is due
to the fact that the magnetic field is presumably anchored in the disk's
midplane, where the pressure is much greater than the atmospheric pressure. 
The magnetic waves propagate upwards along the magnetic flux tube, while 
radiation pressure from the AR is pushing the gas along the magnetic lines, i.e.,
downwards to the disk. This downward direction of the radiation pressure 
arises naturally in a two-phase model (unlike the situation within the 
accretion disk) since here most of the energy is released above the disk's
atmosphere (see also Nayakshin \& Melia 1997b)

With this in mind, we may now describe heuristically how the magnetic
flare develops and how pressure equilibrium is established. As is well 
known (Parker 1979; Galeev et al. 1979), magnetic flux tubes are buoyant 
in a stratified atmosphere, and so they rise to the surface of the accretion 
disk. As the tube is rising, the particles slide along the magnetic field lines
downward to the disk in response to gravity. The magnetic flux tube becomes more
and more particle-free, $v_a$ is increasing, and so the conditions become
more and more favorable for the dissipation of magnetic field energy. 
If magnetohydrodynamic waves are generated and propagate up to the top 
of the flux tube, and if the conditions there for effective reconnection 
are satisfied, the different (incoherent) waves reconnect and thus 
produce highly energetic particles.  The particles in turn produce 
X-radiation by up-scattering the UV radiation from the disk.
Since the radiation pressure $P_r$ is very much smaller than
the stress in the underlying magnetic flux tube, we may neglect the 
sideways expansion of the flux tube. We need to consider the pressure equilibrium
along the magnetic field lines, however, since the plasma can in principle
move freely in that direction.  The balance of radiation pressure with the
magnetic ram pressure then sustains the AR optical depth as discussed above.
Since the flux tube is geometrically thick, the corresponding ratio 
$A_r/A_a$ is probably of order $\sim $ one to a few, and with $v_a\sim c$, we
therefore expect
\begin{equation}
\tau_T\sim 1 - 2\;.
\end{equation}
The lowest values of the equilibrium $\tau_T$ can be reached due to the fact
that 
$A_r$ in this equation is not necessarily the total area of the source,
because some of the X-ray flux can be reflected by the underlying disk
and re-enter the AR. Some of this re-entering flux can be parallel
to the incoming Alfv\'en energy flux, and thus the effective $A_r$ is
smaller than the full geometrical area of the source. Furthermore, we have
assumed a one-temperature gas, and have neglected the gas pressure in 
our calculation. It is possible that the protons are much hotter 
than the electrons, and that they account for a sizable fraction of the total pressure
in the AR, which then leads to a reduction in the value of the equilibrium 
$\tau_T$ as compared to Equation (5).

In the analysis developed here, there is nothing specific
to Alfv\'en waves.  We could have instead invoked the influx
of energy by other waves or even energetic particles 
accelerated by a magnetic reconnection process.  All that matters is that 
they have a well-defined relationship between their momentum and energy
densities, and that their propagation speed is close to $c$. Therefore, 
there may occur a different physical setting where the mechanism
for the pressure equilibrium described here can be at work.
It seems unavoidable that magnetic fields are involved, and that they
therefore provide some of the confinement, e.g., in a direction 
perpendicular to the magnetic field lines. We encourage future work 
on the range of possible geometries for the ARs.

\section{The Range in $\tau_T$ Permitted by Current Observations} 

Zdziarski et al. (1996) produced a fit of the average {\it Ginga}/OSSE 
spectrum of Seyfert 1 galaxies assuming that the active regions form 
hemispheres above the disk.  They found that the radial optical 
depth of the hemispheres is $\tau_T\sim 1$. Here, we will
examine whether the Seyfert spectrum can be due to a combination
of spectral components from flares with different $\tau_T$, but the
same $y$-parameter (set arbitrarily at $1.3$). The latter assumption is
introduced to ensure that the X-ray spectral index does not vary considerably 
from flare to flare. A constant $y$-parameter is a natural consequence of 
the fixed geometry of the flare, in the sense that the cooling
of the plasma is fixed by how much of the X-ray flux re-enters the emitting
region after it is reflected from the disk (see Haardt \& Maraschi 1991).

As an illustration of the method, we first compute the spectrum from flares 
with a range of Thomson optical depths assuming that they all have the same 
luminosity.  We then convolve these spectra with a Gaussian probability
distribution that a flare occurs with $\tau_T$.  The composite
spectrum $F(E)$ (in energy/sec/keV) is 
\begin{equation}
F(E) = \int_{0}^{\infty}\, d\tau_T\; \exp\left [-
{(\tau_T - \tau_0)^2\over \tau_{\sigma}^2}\right ]
F(E, \tau_T)\,{\rm ,}
\end{equation}
where $F(E, \tau_T)$ is the spectrum from a single flare with $\tau_T$.
We take $\tau_0 = 1.14$ and adopt several values of $\tau_{\sigma}={\rm const}$ 
to represent the possible spread in $\tau_T$ between different flares. 
The individual spectra are computed assuming a slab geometry using 
an Eddington frequency-dependent approximation for the radiative transfer, 
using both the isotropic and first moments of the exact Klein-Nishina 
cross section (Nagirner \& Poutanen 1994). Although this geometry is clearly 
different from that of a realistic flare, our point here is
to test the possibility of co-adding spectra with different $\tau_T$, in 
order to see what range in $\tau_T$ may be permitted by current observations.
We expect that a more accurate calculation with the correct geometry will yield
qualitatively similar limits on $\tau_T$, though the exact values should 
be inferred using a $\chi^2$ fit to the data.

Figure 1 shows the results of our calculation for $\tau_0 = 1.14$ 
and two values of $\tau_{\sigma}$: $0.7$ and $1.5$. The spectrum
from a single flare (solid curve) with $\tau_T = 1.14$ is also shown 
for comparison. It can be seen that the plot for $\tau_{\sigma} = 0.7$ is 
hardly distinguishable from that for $\tau_T = 1.14$.  Moreover, these 
curves differ the most above $100$ keV, where the OSSE data typically have 
error bars larger than this difference (see, e.g., Fig. 1 in Zdziarski et al.
1996). On the basis of this simple test, we would expect that
Seyfert spectra may be comprised of contributions from many
ARs encompassing a range ($0.5-2$) of $\tau_T$.  We note, however, 
that a broader range in $\tau_T$ is unlikely because of the considerable
flattening to the spectrum for $\tau_{\sigma}>1.5$.

The conclusion that 
$\tau_T$ is allowed to vary within the range of $0.5-2$ is very important
for the magnetic flare model, since it is otherwise difficult to see how different
flares could produce exactly the same $\tau_T$. It may also happen that
a flare evolves through many phases and that its Thomson optical depth
therefore varies with time.  However, these calculations demonstrate that
as long as that variation is restricted to the range $\sim 0.5-2$, the 
resulting spectrum is consistent with the observations. 

This is not to say that flares with $\tau_T$ greatly in excess of $1$
are not permitted.  According to Equation (5), a large
optical depth probably corresponds to a flare with $v_a\ll c$. 
The magnetic field is therefore probably smaller and the energy is
transported with a smaller $v_a$.  That is, magnetic flares with 
large $\tau_T$ may be significantly dimmer than the more energetic 
ones with $\tau_T\sim 1$, and so they do not contribute noticeably 
to the observed Seyfert spectrum.

\section{Conclusions}

We have considered the consequences of imposing a pressure equilibrium 
on the active regions of Seyfert Galaxies, in addition to the
more often studied energy equilibrium, under the assumption that
the emission arises within energetic magnetic flares
above the surface of a cold disk. We showed that
if the energy is supplied to the X-radiating plasma 
by the influx of some energy source with a group velocity $\sim c$,
then $\tau_T$ probably falls within the range $1-2$. The 
current X-ray/$\gamma$-ray observations are consistent with 
this range of Thomson optical depths. We conclude that magnetic flares 
on the surface of the cold disk remain a viable explanation for the 
spectra observed in Seyfert Galaxies. Alternative explanations,
based on a gravitational confinement of the ARs, appear to be
incapable of accounting for the observed `universality' in 
the value of $\tau_T$.

\section{Acknowledgments}

This work was partially supported by NASA grant NAG 5-3075. We
are very grateful to the anonymous referee for pointing out a
serious error in our original manuscript.


%
%
%

{}

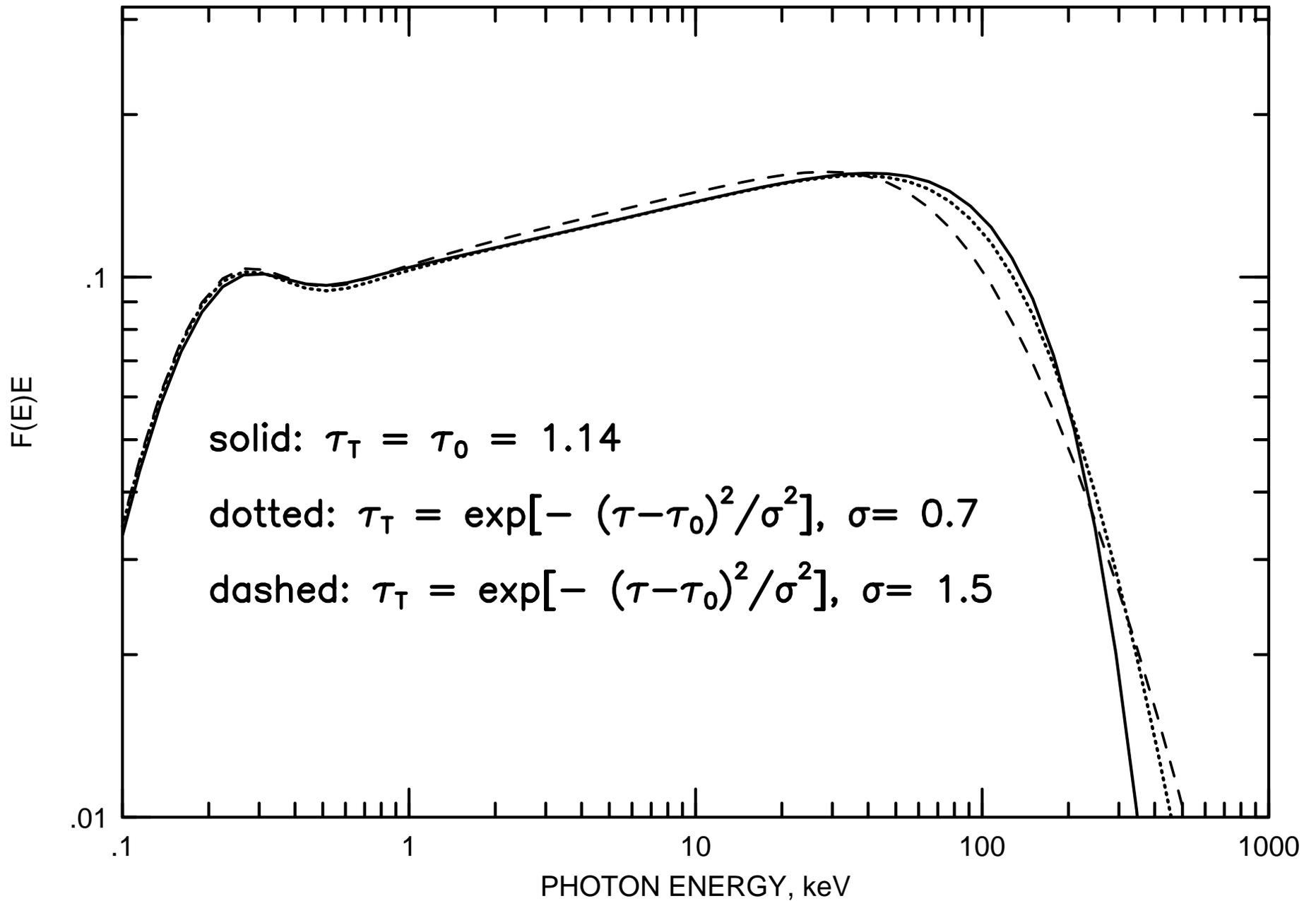
{\bf Figure 1.} The spectrum resulting from co-adding the
components due to different magnetic flares with a Gaussian 
distribution in the Thomson optical depth (see text), 
centered on $\tau_0 = 1.14$ with a width $\tau_{\sigma}=
\, 0.7$ (long-dashed curve) and $1.5$ (short-dashed curve). The
spectrum from a single flare with optical depth $\tau_T =\tau_0$
is shown as a solid curve.  The curves are normalized to the 
same integrated flux.  These spectra demonstrate that current
observations may not be able to differentiate between a single-flare
spectrum and one comprised of many different flares if their 
optical depth is in the range $\sim 0.5 - 2$.

\end{document}